\documentclass[onecolumn,superscriptaddress,floatfix,preprintnumbers,
nofootinbib,hyperref]{revtex4}
\pdfoutput=1
\RequirePackage{fix-cm}
\RequirePackage{graphicx}
\usepackage{amsmath,amssymb}
\usepackage{bm}

\usepackage{xcolor}
\usepackage{braket}

\begin{document}
\title{Neutrino spin oscillations in conformally gravity coupling models and quintessence surrounding a black hole}

\author{Leonardo Mastrototaro}
\email{lmastrototaro@unisa.it}
\affiliation{Dipartimento di Fisica ``E.R Caianiello'', Università degli Studi di Salerno, Via Giovanni Paolo II, 132 - 84084 Fisciano (SA), Italy.}
\affiliation{Istituto Nazionale di Fisica Nucleare - Gruppo Collegato di Salerno - Sezione di Napoli, Via Giovanni Paolo II, 132 - 84084 Fisciano (SA), Italy.}

\author{Gaetano Lambiase}
\email{lambiase@sa.infn.it}
\affiliation{Dipartimento di Fisica ``E.R Caianiello'', Università degli Studi di Salerno, Via Giovanni Paolo II, 132 - 84084 Fisciano (SA), Italy.}
\affiliation{Istituto Nazionale di Fisica Nucleare - Gruppo Collegato di Salerno - Sezione di Napoli, Via Giovanni Paolo II, 132 - 84084 Fisciano (SA), Italy.}

%=============================================================================

\begin{abstract}
In this paper, we study the spin transitions of neutrinos caused by the interaction with a gravitational field. We consider a model with a scalar field (describing screening effects) conformally coupled to matter and neutrinos. The presence of screening effects suppresses the neutrino spin-flip probability as compared with General Relativity predictions. Such a result could be used, combined with neutrino astronomy, for testing modified theories of gravity and, in turn, screening effects invoked to bypass the solar system and Lab tests. Such an analysis has been also extended to the case of the quintessence field surrounding a black hole. Here we investigate the flavor and spin transitions, showing that also in such a case exists a suppression of the effect compared to General Relativity prediction.

  \end{abstract}

\maketitle

\section{Introduction}
\label{intro}

%It is well known that neutrinos %are assumed massless in the %Standard Model, while studies %demonstrate that the masses of at %least two of them are nonzero and %there are mixing between %different neutrino flavors. This %is a direct indication of physics %beyond the standard model. On the %other hand, neutrino mass %introduces some interesting %effects in the coupling with the %gravitational field: one of these %is the spin-flip effect that we %will investigate in this paper.
%
%Moreover, recent cosmological %observations suggest that our %Universe is currently undergoing %to an accelerated expansion %\cite{riess,riess1,riess2,riess3,%riess4,riess5}. To explain such a %phase of the Universe evolution, %physics beyond the standard general relativity is needed  and 
%several {\it alternative} or {\it %modified} theories of gravity %have been proposed. Examples of %alternative theories could be %higher-order curvature %invariants, that allow to get %inflationary behaviour, as well %as to explain the flatness and %horizon problems %\cite{starobinski,starobinski1} %(for further applications and %different models, see Refs. %\cite{Capozziello:2011et,Tino:202%0nla,cosmo1,cosmo2,cosmo3,cosmo4,%cosmo5,cosmo6,cosmo7,cosmo8,cosmo%9,cosmo10,cosmo11,cosmo12,cosmo13%,cosmo14,cosmo15,odi,cosmo17,cosm%o18,cosmo19,cosmo20,cosmo21,cosmo%22,cosmo23,cosmo24,cosmo25,Benett%i:2020hxp}). 

Extended Theories of Gravity (ETG) have been mainly proposed to account for the
recent cosmological observations that suggest that our Universe is currently undergoing to an accelerated expansion \cite{riess,riess1,riess2,riess3,riess4,riess5}.
To account for such a behaviour of the Universe, an unknown form of energy (the Dark Energy) must be necessarily introduced. 
ETG can be obtained in a different way, by generalizing the Hilbert-Einstein action either introducing higher-order curvature invariants, $\mathcal{L}\sim f(R,R_{\mu\nu}R^{\mu\nu},\Box^k R,\dots)$ 
(here $R$ is the Ricci scalar, $R_{\mu\nu}$ the Ricci tensor, and $\Box=\frac{1}{\sqrt{-g}}\partial_\mu (\sqrt{-g}g^{\mu\nu}\partial_\nu)$ the D'Alambertian operator in curved spacetimes, with $g$ the determinant of the metric tensor $g_{\mu\nu}$) or introducing one or more than one scalar fields, obtaining the so-called scalar tensor theories \cite{Capozziello:2011et,Amendola,starobinski,starobinski1} . These generalization of General Relativity are also related to the fact that, at high curvature regimes, curvature invariants are necessary in order to have self-consistent effective actions \cite{birrell,shapiro,barth}.
The ETG allow to address the shortcomings of the Cosmological Standard Model (for example, higher-order curvature invariants allow to get inflationary behaviour, remove the primordial singularity, explain the flatness and horizon problems) \cite{starobinski,starobinski1} (for further applications, see Refs. \cite{Capozziello:2011et,Amendola,Tino:2020nla,cosmo1,cosmo2,cosmo3,cosmo4,cosmo5,cosmo6,cosmo7,cosmo8,cosmo9,cosmo10,cosmo11,cosmo12,cosmo13,cosmo14,cosmo15,cosmo17,cosmo18,cosmo19,cosmo20,cosmo21,cosmo22,cosmo23,cosmo24,cosmo25}). 
Among the various scalar-tensor dark energy models proposed in the literature for explaining the cosmic acceleration of the Universe, it is worth mentioning the model of Refs. \cite{Amendola,[24],[25],[26]}  based on the interaction quintessence-neutrino. Such an interaction is described by a conformal coupling in such a way that when the massive neutrinos became nonrelativistic, by activating the quintessence, the Universe acceleration starts. In these models, the quintessence and matter (including neutrinos) interact, and the interaction is given by a conformal coupling \cite{[28],[29],[30]} that induces in turn screening effects. The latter are mainly classified as chameleon models \cite{[31],[32],[33],[34]} and symmetron models \cite{[35],36a}. The relevant consequence of the screening effects is that the scalar fields behaviour is strongly related to the matter density of the environment, with the consequence that in a dense region they are screened.

This paper aims to investigate the propagation of neutrinos in geometries described by ETG and coupled to quintessence, focusing in particular
on the spin-flip of neutrinos when they scatter off Black Holes (BH).
Experiments on neutrino physics provide clear evidence that neutrino oscillates in different flavors \cite{Ace19,Ker20,Aga19}.
These results not only give indirect proof of the fact that neutrinos are massive particles but, in turn,
they represent an indication of physics beyond the Standard Model.
New possibilities to study the neutrino properties are offered by the interactions with external fields,
which could be magnetic fields or gravitational fields, to which we are interested in.
In the first case, the formulas of the oscillation probabilities of neutrino in different flavors are affected if
neutrinos interact with external fields \cite{BalKay18}, as well as interactions with electromagnetic fields may also induce a helicity transition
of neutrinos with different helicities. These processes are generically called spin oscillation and/or spin-flavor oscillations~\cite{Giu19}.
The latter is also influenced when neutrino propagates in a curved background.
It is well known that the gravitational interaction can affect the neutrino oscillations or induce the change of the polarization
of a spinning particle \cite{Pap51,PirRoyWud96,SorZil07,ObuSilTer17,Dvo06,Dvo19,Cuesta,luca,Cardall,Visinelli,Chakraborty,Sor12,Ahluwalia,Punzi,Swami,CuestaApJ,LambMNRAS,Capozz}.

Here we study the helicity transitions (spin oscillations) of neutrinos $\nu_{fL}\to\nu_{fR}$, in which
neutrino flavors do not change under the influence of external gravitational fields (the case of neutrino flavor oscillations has been studied in \cite{Sadjadi:2020ozc}).
Since in the Standard Model neutrinos are produced with fixed left-handed polarization, a change in right-handed polarization
induced by a gravitational field would mean that they become sterile, and therefore do not interact (except gravitationally). As a consequence,
a detector would register a different neutrino flux, giving a signature of the coupling of neutrinos with quintessence fields screening the gravitational source.

We also discuss the effects of the quintessence field surrounding a black hole on neutrino flavor oscillations and neutrino spin flip. The existence of a quintessence diffuse in the Universe has opened the possibility that it could be present around a massive gravitational object, deforming the spacetime around a gravitational source. In \cite{Kiselev:2002dx} the Einstein field equations have been solved for static spherically symmetric quintessence surrounding a black hole in $d=4$ dimensions. As a result, the Schwarzschild geometry gets modified.  

The paper is organized as follows. In the next Section, we review the spin-flip transition in a general curved spacetime.
In Section 3 we consider neutrino interaction through the conformal coupling \cite{[28],[29],[30]} responsible for screening effects, and compute the transition probabilities. In Section 4 we study the neutrino flavor and spin transition in a background described by a black hole surrounding by a quintessence field. We shortly analyzed the quintessence field on nucleosynthesis processes.
In the last Section, we discuss our conclusions.

\section{Neutrino spin evolution in a generic gravitational field}
\label{Neutrino spin evolution in a generic}

In this Section we  treat the neutrino spin oscillation problem in a generic gravitational metric. We follow the papers by Dvornikov \cite{Dvornikov:2020oay} and Obukhov-Silenko-Teryaev \cite{Obukhov:2009qs}. The motion of a spinning particle in gravitational fields is related to its spin tensor $S^{\mu\nu}$ and momentum $p^{\mu}$ 
\begin{align}
    \frac{DS^{\mu\nu}}{D\lambda}&=p^{\mu}v^{\nu}-p^{\nu}v^{\mu} \,\ \\
    \frac{Dp^{\mu}}{D\lambda}&=-\frac{1}{2}R^{\mu}_{\nu\rho\sigma}v^{\nu}S^{\rho\sigma} \,,
\end{align}
where $v^{\mu}$ is the unit tangent vector to the center of mass world line, $\lambda$ is the parameter, $D/D\lambda$ is the covariant derivative along the world line and $R^{\mu}_{\nu\rho\sigma}$ is the Riemann tensor. One can define the spin vector as
\begin{equation}
    S_{\rho}=\frac{1}{2m}\sqrt{-g}\epsilon_{\mu\nu\lambda\rho}p^{\mu}S^{\nu\lambda} \,\ ,
\end{equation}
with $\epsilon_{\mu\nu\lambda\rho}$ the completely antisymmetric tensor, $g$ the determinant of the metric $g_{\mu\nu}$ and $m^2=p_{\mu}p^{\mu}$.
Using the principle of General Covariance, the particle motion has to satisfy the following relations
\begin{align}
    \frac{DS^{\mu}}{D\tau}&=0 \,, \\
    \frac{DU^{\mu}}{D\tau}&=0 \,.
\end{align}
where $U^{\mu}=dx^{\mu}/d\tau$ and $\tau$ is the proper time. This means that
\begin{align}
    \frac{dS^{\mu}}{d\tau}=-\Gamma^{\mu}_{\alpha\beta}U^{\alpha}S^{\beta} \,, \\
    \frac{dU^{\mu}}{d\tau}=-\Gamma^{\mu}_{\alpha\beta}U^{\alpha}U^{\beta} \,.
\end{align}
However, in the particle description, what is relevant is the spin measured in the rest frame of the particle: we will use the tetrads $V^{a}_{\mu}$ to do the transformation. They are defined as
\begin{equation}
    g_{\mu\nu}=V^a_{\mu}V^b_{\nu}\eta_{ab} \,,
\end{equation}
where $\eta_{ab}$ is the Minkowski metric. The equations in the local frame have the form:
\begin{align}
    \frac{ds^a}{dt}=\frac{1}{\gamma}G^{ab}s_b \,, \\
     \frac{du^a}{dt}=\frac{1}{\gamma}G^{ab}u_b \,,
\end{align}
where $s^a=S^{\mu}V^a_{\mu}$, $u^a=U^{\mu}V^a_{\mu}$, $\gamma=U^0=dt/d\tau$, $G^{ab}=\eta^{ac}\eta^{bd}\gamma_{cde}u^e$, $\gamma_{abc}=\eta_{ad}V^{d}_{\mu;\nu}V_b^{\mu}V_c^{\nu}$ and
\begin{equation}
    V^{d}_{\mu;\nu}=\frac{DV^d_{\mu}}{dx^{\nu}} \,.
\end{equation}
The evolution of the spin vector $\mathbf{s}^a$ is given by
\begin{equation}
    \frac{d\mathbf{s}^a}{dt}=\frac{2}{\gamma}(\bm{\zeta}\times \mathbf{G})=2\,\bm{\zeta}\times\mathbf{\Omega_g} \,,
    \label{evolution}
\end{equation}
where $\mathbf{\zeta}$ and $\mathbf{G}$ are defined as
\begin{align}
    \mathbf{s}^a&=\left(\bm{\zeta\cdot u},\mathbf{\zeta}+\frac{\bm{u(\zeta \cdot u)}}{1+u^0}\right) \,, \\
    \mathbf{G}&=\frac{1}{2}\left(\mathbf{B}+\frac{\mathbf{E}\times\mathbf{u}}{1+u^0}\right) \,, \\
    u&=(u^0,\mathbf{u}) \,,
\end{align}
with $G_{0i}=E_i$ and $G_{ij}=-\epsilon_{ijk}B_k$.
In the metric of our interest, $\Omega_g=(0,\Omega_2,0)$ and therefore we can use the following representation for $\bm{\zeta}=(\zeta_1,0,\zeta_3)=(\cos{\alpha},0,\sin{\alpha})$.

We are interested into studying the neutrino spin oscillation and therefore we focus on the helicity of the particle $h=\bm{\zeta \cdot u}/|\bm{u}|$. The initial helicity for a  neutrino is $h_{-\infty}=-1$ and defining the initial condition $\bm{u}_{-\infty}=\left(-\sqrt{E^2-m^2},0,0\right)$ one can get that $\bm{\zeta}_{-\infty}=(1,0,0)$, $\alpha_{-\infty}=0$. Moreover, we can write that $\bm{u}_{+\infty}=\break \left(+\sqrt{E^2-m^2},0,0\right)$ and therefore $h_{+\infty}=\cos{\alpha}$.  

From that, we can state that the helicity states of the neutrino can be written as:
\begin{align}
    \psi_{-\infty}&=\ket{-1} \,\ \\
    \psi_{+\infty}&=a_+\ket{-1}+a_-\ket{1} \,,
\end{align}
where $a_+^2+a_-^2=1$ due to the normalization and $a_+^2-a_-^2=\cos\alpha=\langle h\rangle_{+\infty}$. From that, one obtains that $a_{\pm}^2=(1\pm\cos{\alpha})/2$ and the probability to find a neutrino with right-handed helicity is
\begin{equation}\label{PLR18}
P_{LR}=|a_-|^2=\frac{1-\cos\alpha_{+\infty}}{2} \,.
\end{equation}
Using Eq. (\ref{evolution}), we obtain
\begin{equation}
    \frac{d\sin\alpha}{dt}=2\cos\alpha\Omega_2\quad\rightarrow\quad\alpha=2\Omega_2 t \,.
\end{equation}
Therefore, it is possible to write
\begin{equation}
    \frac{d\alpha}{dr}=\frac{d\alpha}{dt}\frac{dt}{dr}=\frac{d\alpha}{dt}\frac{dt}{d\tau}\frac{d\tau}{dr} \,,
\end{equation}
where $dt/d\tau=U^0$ and $dr/d\tau=U^1$. Finally, the angle $\alpha_{+\infty}$ reads
\begin{equation}
    \alpha_{+\infty}=\int dr\frac{d\alpha}{dr} \,\ .
    \label{21}
\end{equation}

\section{Neutrino propagating  in conformal metric (screening effects)}

We consider the action with a scalar field conformally coupled to matter 
\begin{eqnarray}\label{Action1}
S= \int d^4x \sqrt{-g}\Bigg[\frac{M_{p}^2}{2}R - \frac{1}{2} g^{\mu\nu} \partial_\mu \phi \partial_\nu \phi - V(\phi) \Bigg] +   
+ \int d^4x  {\cal L}_m \left(\Psi_i , \tilde{g}_{\mu\nu}\right),
\end{eqnarray}
where ${\cal L}_m$ is the Lagrangian density of the matter fields $\Psi_i$, the metric $\tilde{g}_{\mu\nu}$ is related to the metric $g_{\mu\nu}$, by the relation \cite{Faraoni:1998qx,Carneiro:2004rt,Bean}
\begin{equation}\label{gtilde}
\tilde{g}_{\mu\nu} = A^2(\phi) g_{\mu\nu}\,,
\end{equation}
and $M_{p}$ is the reduced Planck mass.
The conformal factor $A(\phi)$ appearing into Eq. (\ref{gtilde}) is a function of the scalar field $\phi$ and induces screening effects. Different screening mechanisms have been proposed in literature, such as, the chameleon \cite{[31],[32],[33],[34]} and symmetron \cite{[35],36a} mechanisms (see also the Vainshtein mechanism \cite{Vainshtein}). They differentiate by the different choices of the coupling and potential functions.
For example, in the case of chameleon mechanism one chooses 
\begin{equation}\label{Aphi}
A(\phi) \equiv \exp\left[{\frac{1}{M_p} \int \beta(\phi) d\phi}\right],
\end{equation}
where $\phi$ is the chameleon conformal field, and $\beta(\phi)$ is a field-dependent coupling parameter (a simple choice is $\beta$ a constant value of the order $\beta \sim {\cal O}(1)$). In the case of the symmetron mechanism, the $\mathbb{Z}_2$-symmetry imposes to choose a quadratic coupling function, $A(\phi) \equiv 1 + \frac{\phi^2(r)}{2M^2}$. In what follows we shall assume 
that $A(\phi)$ is universal in order to respect the equivalence principle.

%\subsection{Spin-flavor in conformal coupling gravity}

%\section{Spin-flip in conformal coupling gravity}

\begin{figure}
    \centering
    \includegraphics[scale=0.9]{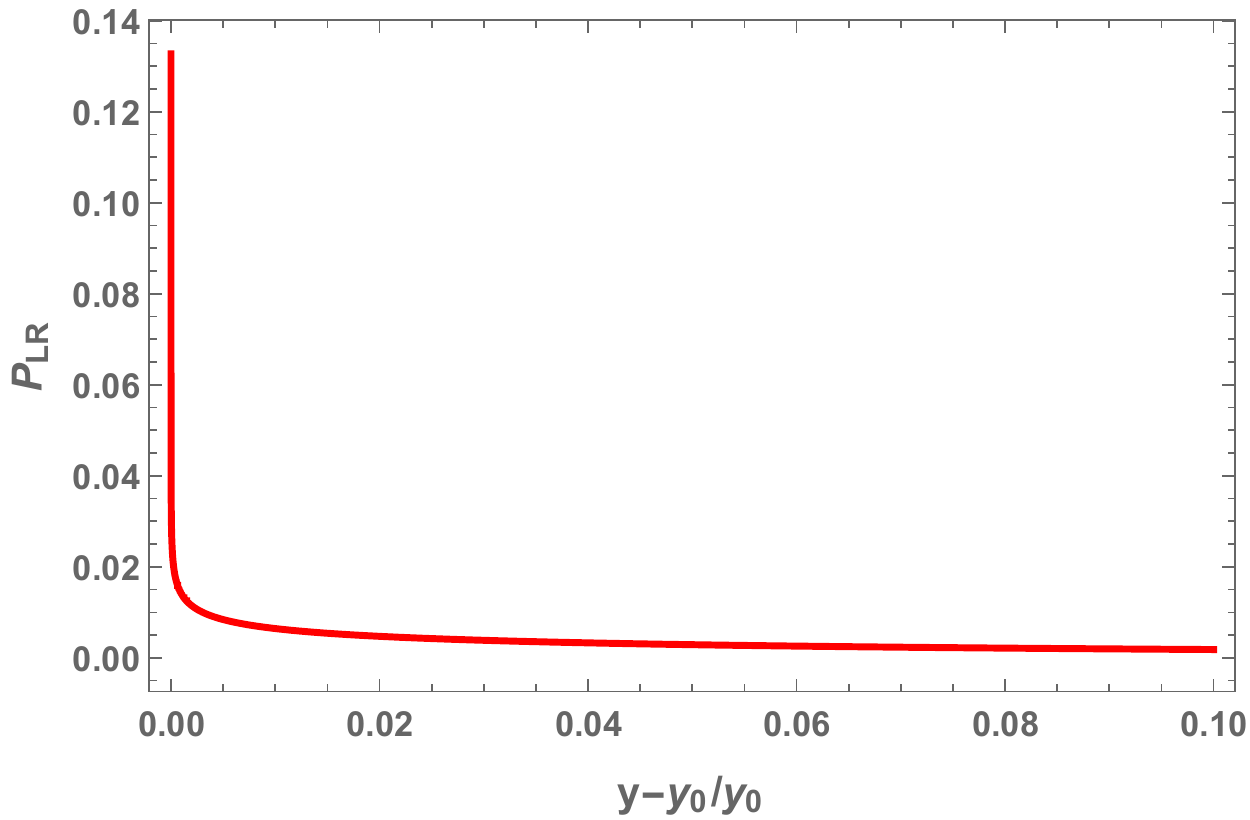}
    \caption{Probability of spin flip plotted respect to the value of the variable $y$. As it can be see, for larger value of $y$ the probability goes to zero. We have used the value of $\gamma=10$ and $A=1$.}
    \label{Conf1}
\end{figure}

\begin{figure}
    \centering
    \includegraphics[scale=0.9]{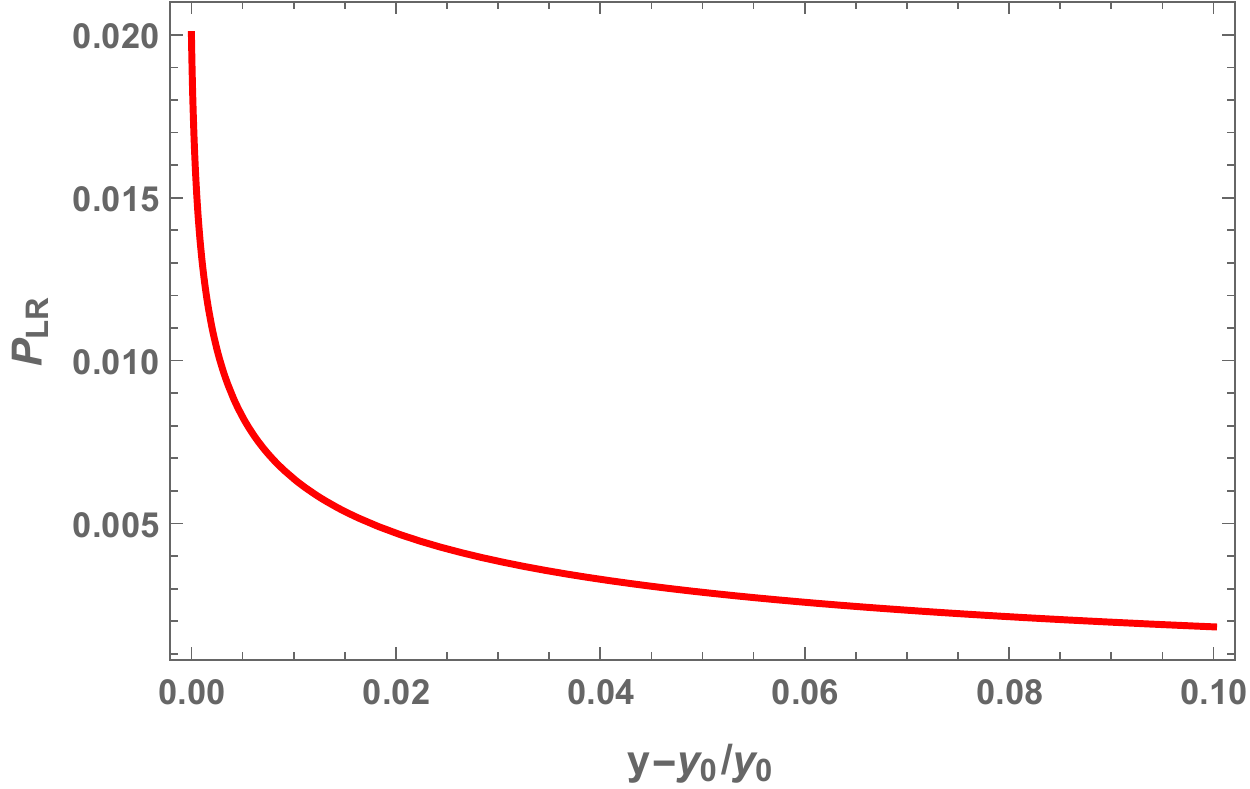}
    \caption{Probability of spin flip plotted respect to the value of the variable $y$. As it can be see, for larger value of $y$ the probability goes to zero. We have used the value of $\gamma=10$ and $A=1.0005$.}
    \label{Conf1.0005}
\end{figure}

\begin{figure}
    \centering
    \includegraphics[scale=0.9]{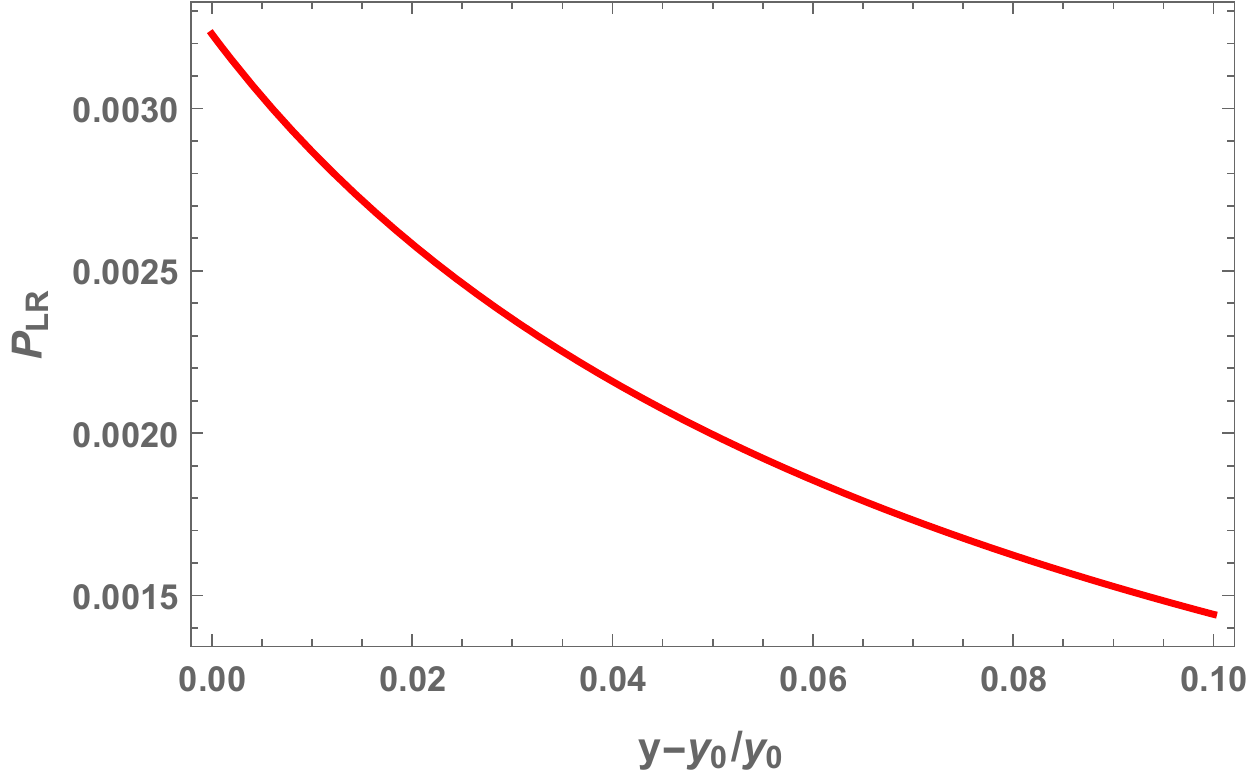}
    \caption{Probability of spin flip plotted respect to the value of the variable $y$. As it can be see, for larger value of $y$ the probability goes to zero. We have used the value of $\gamma=10$ and $A=1.1$.}
    \label{Conf1.1}
\end{figure}

To apply the results of Sec.~\ref{Neutrino spin evolution in a generic}, we consider neutrino propagation near a non-rotational BH in a conformal metric (see (\ref{gtilde})), with $g_{\mu\nu}$ given by 
\begin{equation}
    g_{\mu\nu}=A(\phi)\left(-f(r),\frac{1}{f(r)},r^2,r^2\sin^2\theta\right) \,,
\end{equation}
with $f(r)=1-2M/r$ (Schwarzschild geometry) . Since the metric is spherically symmetric, we can take the motion of the neutrino in the equatorial plane ($\theta=\pi/2$ and $d\theta=0$). With this condition, we obtain
\begin{equation}
    \Omega_2=\frac{L\sqrt{1-\frac{2M}{r}}}{2Er^2}\frac{Am\left(1-\frac{2M}{r}\right)^{\frac{3}{2}}+E\left(1-\frac{3M}{r}\right)}{Am\left(1-\frac{2M}{r}\right)+E} \,.
\end{equation}
From Eq.~(\ref{21}) we can define
\begin{equation}
    \alpha_{+\infty}=\int_{x_m}^{+\infty}\frac{d\alpha}{dx}\frac{dx}{dr}dr \,,
\end{equation}
where $x_m$ is the minimumm value of $x=r/2M$ allowed in the expression of $d\alpha/dx$. Indeed, it is possible to write
%%%%%%%%%%%%%%%%%
\begin{eqnarray}
\frac{d\alpha}{dr}&=&
    \frac{4L}{m}\sqrt{1-\frac{2M}{r}} \frac{\cal A}{\cal B}\,, \\
    {\cal A}&\equiv &
    A\left(-2+\frac{r}{M}\right)^2+\frac{Er}{mM}\left(\frac{r}{M}-3\right)\sqrt{1-\frac{2M}{r}}\,, \nonumber \\
    {\cal B}  &\equiv& 
    \sqrt{\frac{\left(-1+\frac{E^2}{m^2A(1-2M/r)}-\frac{L^2M^2}{Am^2r^2}\right)\left(1-\frac{2M}{r}\right)}{A}} \left(\sqrt{A}\left(-2+\frac{r}{M}\right)+\frac{Er}{mM}\sqrt{1-\frac{2M}{r}}\right) \frac{Ar^3}{M^3} \Bigg(-1 + \frac{2M}{r}\Bigg)\,, \nonumber 
\end{eqnarray}
%%%%%%%%%%%%%%%%%
%
%\begin{equation}
%\begin{split}
%    &\frac{d\alpha}{dr}=\\
%%    &\frac{4L}{m}\sqrt{1-\frac{2M}{r}}\%left(A(-2M+r)^2+\frac{Er}{m}(-3M+r)\sqrt%%{1-\frac{2M}{r}}\right)\\
%    &\Bigg(A\sqrt{\frac{\left(-1+\frac{E%^2}{m^2A(1-2M/r)}-\frac{L^2}{Ar^2}\right%%)\left(1-2M/r)\right)}{A}}\Bigg(-1\\
%    &+\frac{2M}{r}\Bigg)r^3\left(\sqrt{A%}(-2M+r)+\frac{E}{m}\sqrt{1-\frac{2M}{r}%}r\right)\Bigg)^{-1} \,\ ,
%    \end{split}
%\end{equation}
%
where the $x_m$ is given by the condition
\begin{equation}
   \left(\frac{E^2}{m^2A(1-2M/r)}-\frac{L^2M^2}{Am^2r^2}-1\right)\frac{r-2M}{Ar}>0 \,\ . 
\end{equation}
Moreover, it is useful to use the following variable
\begin{align}
    y&=\frac{b}{2M} \,,\\
    b&=\frac{L}{E}\frac{1}{\sqrt{1-\gamma^{-2}}} \,, \\
    \gamma&=\frac{E}{m} \,,
\end{align}
where $y>y_0$ with $y_0$ the critical impact parameter that for massive particle depends on $\gamma$. The critical impact parameter can be found from the effective potential of black hole
\begin{equation}
    V_{\mathrm{eff}}=\frac{Am^2(1-\frac{1}{\gamma^{2}})}{L^2}\left(1-\frac{2M}{r}\right)\left(1+\frac{L^2}{Am^2r^2}\right) \,,
    \end{equation}
imposing $dV_{\mathrm{eff}}/dr=0$, finding the maximum and then solve the equation
\begin{equation}
    \left(\frac{dr}{d\tau}\right)^2=0=\frac{1}{b^2}-V_{\mathrm{eff}} \,.
    \label{bcrit}
\end{equation}
From Eq.~(\ref{bcrit}), one can obtain the critical impact parameter. Finally one can use Eq.~(\ref{21}) to find the probability of a neutrino spin flip. Results, for $\gamma=10$, are shown in Fig.~\ref{Conf1},\ref{Conf1.0005} and \ref{Conf1.1}.
In Fig.~\ref{Conf1}, we have used a factor $A=1$, recovering the results in Ref.~\cite{Dvornikov:2020oay} (with a factor $1/2$ overall). 
In the case of the chameleon theory, the conformal factor $A(\phi)$ is given by (\ref{Aphi}). The simplest conformal model is that with $\beta$ constant and $A=\exp(\beta(\phi)\phi/M_p)$. Referring to \cite{Sadjadi:2020ozc} (see Fig. 10 and 11), it turns out that $(\beta,\phi)$ may assume the values $(1,1.483\times 10^{27}~\mathrm{eV})$ or $(10,4.69\times 10^{26}~\mathrm{eV})$ which lead to $A\geq 1.1$. However, in general, using the exponential definition it results that $A\geq 1$.
As shown in Fig. \ref{Conf1.1}, the neutrino spin oscillation probability in chameleon theories is suppressed respect to GR, so that the flux of (non-relativistic) neutrinos with initial fixed (left-handed) polarization arriving at detector remains unaltered. 
It is worth noting that also small deviations from $A=1$ (GR case) induce suppression of the spin-flip probability, as shown in Fig.~\ref{Conf1.0005} for $A=1.0005$.

%we analyze the case $A=1.0005$ and %$A=1.1$, respectively, to show that a %small changing in the conformal factor %can affect in a non-negligible way the %results even by two orders of magnitude %leading to an effect that could probe %even small modification of the GR %metric.

\section{Neutrinos interacting with Quintessence field}

As a second example, we consider neutrinos propagating near a non-rotational black hole surrounding by a  quintessence field \cite{Kiselev:2002dx,Chen:2008ra}
\begin{equation}
    g_{\mu\nu}=\left(-f_1(r),\frac{1}{f_1(r)},r^2,r^2\sin^2\theta\right) \,\ ,
\end{equation}
where $f_1(r)=1-2M/r-c/r^{3\omega_q+1}$, where $c$ is a positive constant and $-1<\omega_q<-1/3$.

\subsection{Neutrino flavor oscillations}

Let us shortly recall the neutrino oscillations in curved space-times (see, for example, Refs.  \cite{1997Fornengo,Cardall}, and references therein)
Neutrino flavor oscillations occur owing the fact that neutrino flavor eigenstates $|\nu_\alpha\rangle$ are linear combinations of neutrino mass eigenstates $|\nu_j\rangle$
as
\begin{equation}
|\nu_{\alpha}\rangle= \sum_i U_{\alpha i}\, e^{-i\Phi_i} |\nu_{j}\rangle ,
\label{neutreigenst}
\end{equation}
where $\alpha$ ($i$) labels the neutrino flavor (mass) eigenstates, while $U_{\alpha j}$ is the (unitary) mixing  matrix between the flavor eigenstates and the mass eigenstates. The phase  $\Phi_j$ is associated to the $i$\textit{th} mass eigenstate, and in a curved spacetimes  reads
\begin{equation}
\Phi_i= \int P_{(i)\mu}dx^{\mu}
\label{phaseosc}\,.
\end{equation}

Here $P_{(i)\mu}$ indicates the four-momentum of the mass eigenstate $i$. In what follows we shall assume that neutrinos just have two flavors, so that introducing the mixing angle $\Theta$, the transition probability from one flavor eigenstate $\alpha$ to
another $\beta$ is given by
\begin{equation}
P(\nu_{\alpha}\rightarrow \nu_{\beta})= \sin^2(2\Theta)\sin^2\left(\frac{\Phi_{jk}}{2}
\right),
\label{transprob}
\end{equation}
where $\Phi_{jk}\doteq \Phi_j-\Phi_k$.
The explicit form of $\Phi_i$ in a Schwarzschild like geoemtry reads  \cite{1997Fornengo,Cardall,Cuesta}
\begin{eqnarray}\label{phaseosci}
\Phi_j &=& \int dr\frac{m_j}{\dot{r}}=\int \frac{  m_j^2 dr}{\sqrt{E^2 -
g_{00}(r) \left[\frac{L^2}{r^2}+
m_j^2 \right]}} \,\ , 
\end{eqnarray}
where $L$ represents the angular momentum of particles. Equation (\ref{phaseosci}) is exact. The phase $\Phi_i$ vanishes for null geodesics\footnote{This follows by the fact that $p_{\mu}dx^{\mu}=g_{\mu\beta}p^{\beta}dx^{\mu}\propto ds^2$,
which vanishes for null paths}.

\begin{figure}
    \centering
    \includegraphics[scale=0.9]{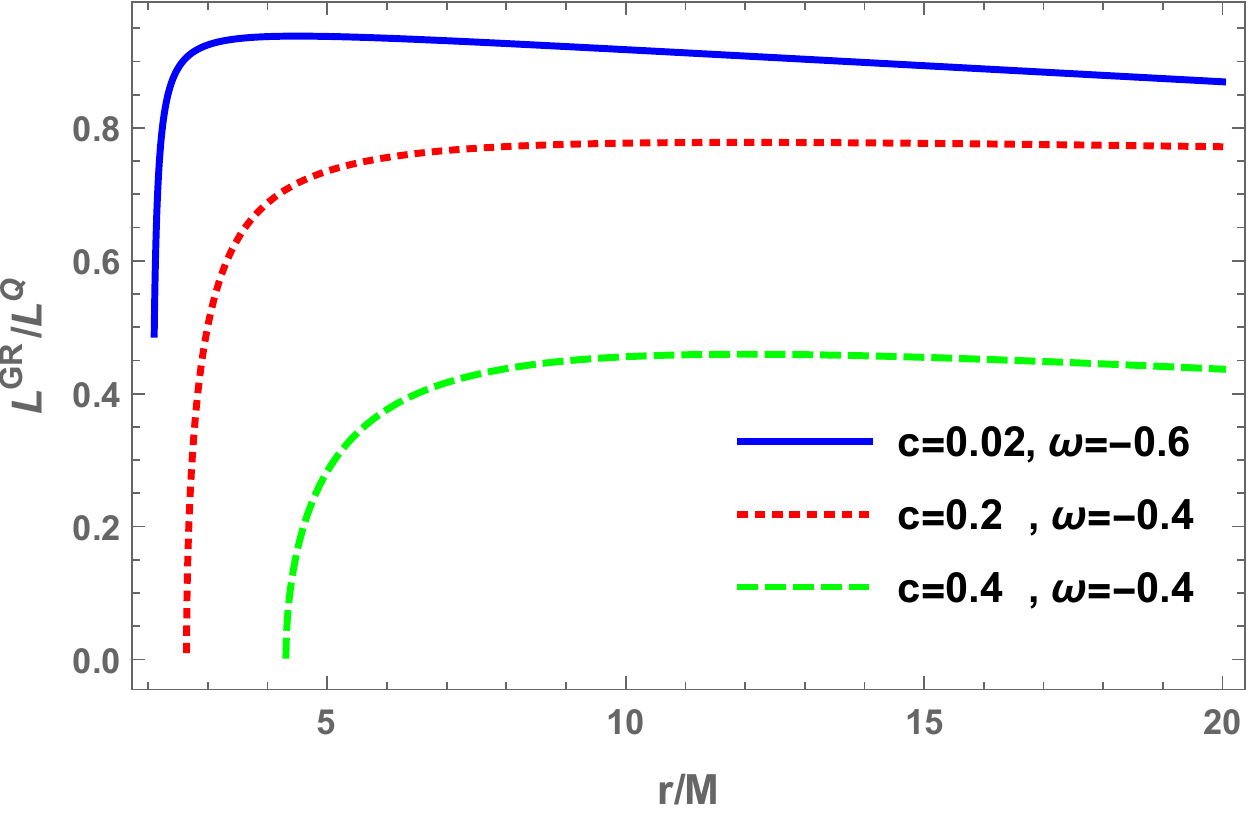}
    \caption{Oscillation length ratio between GR and the quintessence framework in function of $r/M$. The legend shows the values of $c$ and $\omega$ used.}
    \label{Quint_OL2}
\end{figure}

From \cite{Cuesta} one infers the neutrino oscillation length, which estimates the length over which a given neutrino has to travel for $\Phi_{jk}$ to change by $2\pi$.
Assuming that the particles involved have the same energy $E$, with $E\gg m_{j,k}$, the oscillation length is given by
\begin{equation}
L_{osc}\doteq \frac{dl_{pr}}{d\Phi_{jk}/(2\pi)} \simeq \frac{2\pi
E}{\sqrt{g_{00}} (m^2_j-m^2_k)}\,,
\label{osclength}
\end{equation}
where $dl_{pr}=\left(-g_{ij}\right)dx^idx^j$ is the infinitesimal proper distance (for a Schwarzschild-like geometry),
with $i,j=1,2,3$ (here we are using the natural units -  the conventional units are restored multiplying the right-hand side of (\ref{osclength}) with $\hbar/c^3$). As we can see from  (\ref{osclength}), the oscillation length decreases whenever $g_{00}$ increases.
%This is exactly the case for nonlinear charged black holes, when
%compared to a Schwarzschild black hole. This means that when the black hole is charged,
%neutrinos will tend to oscillate more than they would when it is not charged, for each spacetime point (location).
In Fig.~\ref{Quint_OL2} it is possible to see the ratio between the oscillation length in GR and that in the quintessence framework with $\omega=-0.4,c=0.4$, $\omega=-0.4,c=0.2$ and $\omega=-0.6,c=0.02$ respectively. As it can be seen, increasing $c$ with the same $\omega_q$ or lowering $\omega_q$ with the same $c$ tends to enhance the difference between GR and the quintessence metric.

\subsection{Spin-flip transition}

\begin{figure}
    \centering
    \includegraphics[scale=0.9]{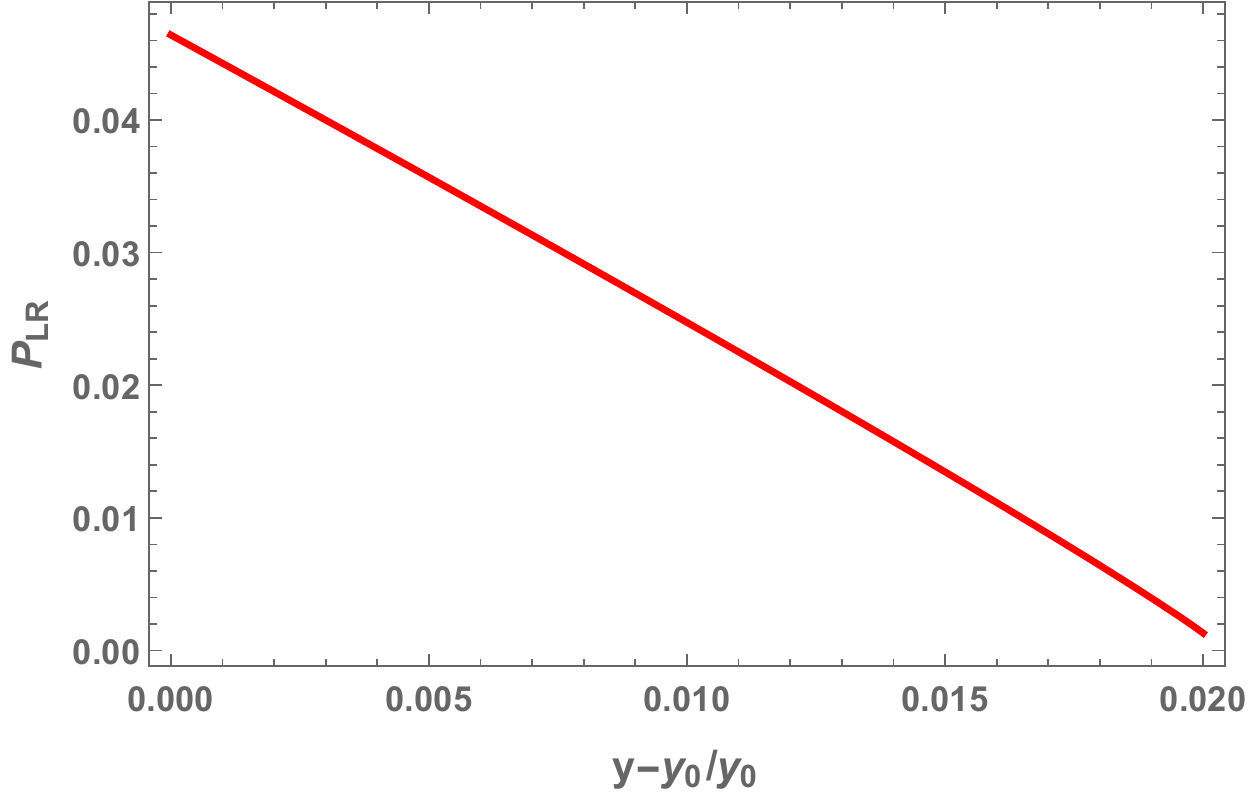}
    \caption{Probability of spin flip plotted respect to the value of the variable $y$. As it can be see, for larger value of $y$ the probability goes to zero. We have used the value of $\gamma=10$, $c=0.4$ and $\omega_q=-0.4$.}
    \label{Quint}
\end{figure}

With a similar procedure as in Sec.~3, we obtain the $\Omega_2$ value that we do not report here to its complex analytic form. To account for the maximum possible difference in the neutrino oscillation with quintessence, we have done a theoretical analysis with $\omega_q=-0.4$ and $c=0.4$. Moreover, due to the complexity of calculation, we can only obtain a numerical approximated result that is represented in Fig.~\ref{Quint}.

As it can be seen, even in this case, the probability is suppressed with respect to the probability computed with the GR metric.

\subsection{Electron fraction $Y_{e}$ in presence of quintessence }

In this Subsection we discuss the effects of gravity (gravitational redshift) on the energy spectra of neutrinos (${\nu}_e$) and antineutrinos ($\overline{\nu}_e$) 
outflowing from the very inner ejecta of a Type II supernova explosion \cite{1996NuPhA.606..167F}. 
The ${\nu}_e-\overline{\nu}_e$ oscillations mediated by the
gravitational collapse of the supernova inner core could explain the abundance of
neutrons. This in turn affects the r-process nucleosynthesis in astrophysical environments\footnote{If indeed $\overline{\nu}_e$ could be over-abundant than $\nu_e$, then the neutron production could be higher than the proton production, and the supernova spin-flip conversion ${\nu}_e \longrightarrow \overline{\nu}_e$ (for example, Majorana type neutrinos) could be affected by gravity-induced effects inside supernovae cores, and hence, the over-abundance of neutrons required for the r-process in such a spacetime with quintessence.}.

\begin{figure}
    \centering
    \includegraphics[scale=0.9]{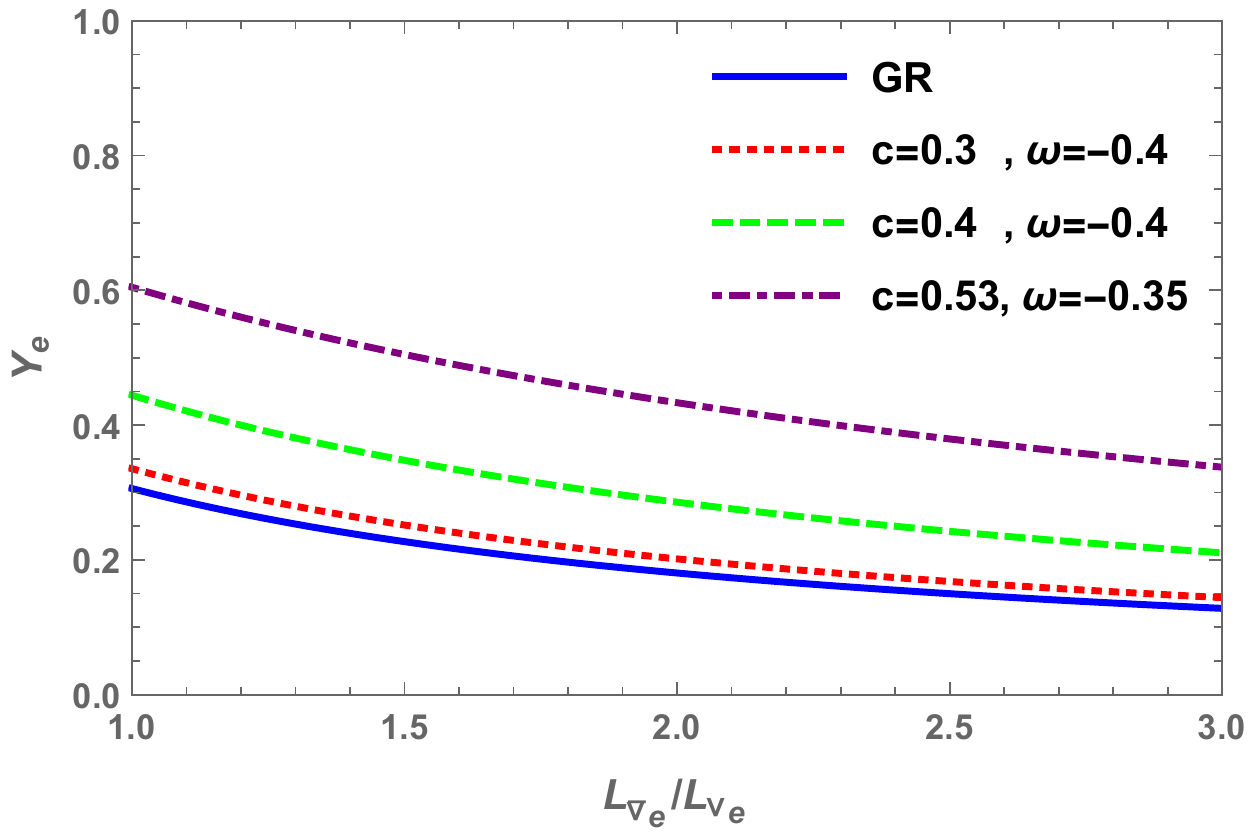}
    \caption{Electron fraction $Y_e$ vs the antineutrino/neutrino luminosity ratio $ L_{\overline{\nu}_e } / L_{\nu_e}$ for $r=5M$. We compare the Schwarzschild electron fraction (GR) with the ones coming from the quintessence theory with the parameters in the legend.}
    \label{Ye}
\end{figure}

By defining ${\nu}_e$ the neutrinosphere at $r_{\nu_e}$ and the
$\overline{\nu}_e$ neutrinosphere at $r_{\overline{\nu}_e} $, the electron fraction reads \cite{1996NuPhA.606..167F} (for details, see also \cite{APJL789-M.Shibata-2014,Cuesta})
\begin{equation}
Y_e = \frac{1}{ 1 + R_{\frac{n}{p}} }, \hskip 1.0truecm R_{\frac{n}{p}} \equiv
R^0_{\frac{n}{p}} \, \Gamma,
\label{gravitYe}
\end{equation}
where $R^0_{\frac{n}{p}}$ (the local neutron-to-proton ratio) and $\Gamma$ are given by
\begin{equation}%\label{neurt-prot-ratio}
R^0_{\frac{n}{p}} \simeq \left[\frac{ L_{\overline{\nu}_e } \, \langle
E_{\overline{\nu}_e} \rangle}{ L_{\nu_e} \, \langle
E_{\nu_e}\rangle} \right]\,, \quad  \Gamma \equiv \left[\frac{g_{00}(r_{\overline{\nu}_e})}{g_{00}(r_{{\nu}_e})}\right]^{\frac{3}{2}}\,, %\label{Gamma}\,,
\label{R0np}
\end{equation}
while $L_{\overline{\nu}_e, \nu_e}$ is the neutrino luminosity and 
$\langle E_{\overline{\nu}_e, \nu_e}\rangle$  is the average energy\footnote{Notice that it is assumed that the ${\overline{\nu}_e}, {\nu}_e$ energy spectrum does not evolve significantly with increasing radius above the ${\overline{\nu}_e}, {\nu}_e$ sphere, as a consequence of the concomitant emission, absorption
and scattering processes.}  (as measured by a
locally inertial observer at rest at the $\{\overline{\nu}_e, \nu_e\}$-neutrinosphere). 
In Fig. \ref{Ye} we plot the electron fraction given by (\ref{gravitYe}).
The presence of quintessence field surrounding a black hole change the neutron-to-proton ratio with respect to the Schwarzschild case ($Y_{e} > Y_{e}^{GR}$), favoring in such a case  the r-processes.\\
On the other hand, there are not relevant differences in $Y_e$ between conformal theory and GR due to the fact that in Eq.~(\ref{R0np}) the conformal factor is almost totally simplified.

%\subsection{limit of 0.02}
%Following the treatment in Ref.~\cite{Chakraborty}, it is possible to obtain the probabily of a spin-flip alon  a radial geodetic:
%\begin{equation}
%    P=\sin^2\left(\frac{Lf(r)\sqrt{f(%r)}}{2Er^2}\left(1-\frac{f'(r)r}{2f(r%)}\right)r\right) \,\ ,
%\end{equation}
%where $f'(r)$ is the derivative of %$f(r)$ in r. This equation gives a %probability under the constraint of %$0.02$ for both the metrics. %\textcolor{
%red}{sicuri vogliamo aggiungerlo?}

\section{Conclusions}

In this paper, we have analyzed the neutrino spin-flip and spin-flavor phenomena in a gravitational field. 
After discussing the general results we have applied them to two specific cases: 1) the conformal modification of GR, which is an effect purely geometrical, related to screening effects, and 2) the quintessence surrounding a black hole, which is linked to dark matter and energy. Regarding the first case, as discussed in the paper, they are characterized by the introduction of an additional degree of freedom (typically a scalar field) that obeys a non-linear equation that couples to the environment.
Screening mechanisms allow circumventing Solar system and laboratory tests by suppressing, in a dynamical way, deviations from GR 
(the effects of the additional degrees of freedom are hidden, in high-density regions, by the coupling of the field with matter
while, in low-density regions, they are unsuppressed on cosmological scales \cite{veltman,symmetron,Vainshtein}). 
Therefore, new tests of the gravitational interaction may provide a new test for probing the existence of these scalar fields.
The gravitational interaction described by {\it deformed} Schwarzschild's geometry, induced hence by the presence of scalar fields around the gravitational massive source, affects the neutrino flavor and spin-flip transitions. 

\begin{figure}
    \centering
    \includegraphics[scale=0.6]{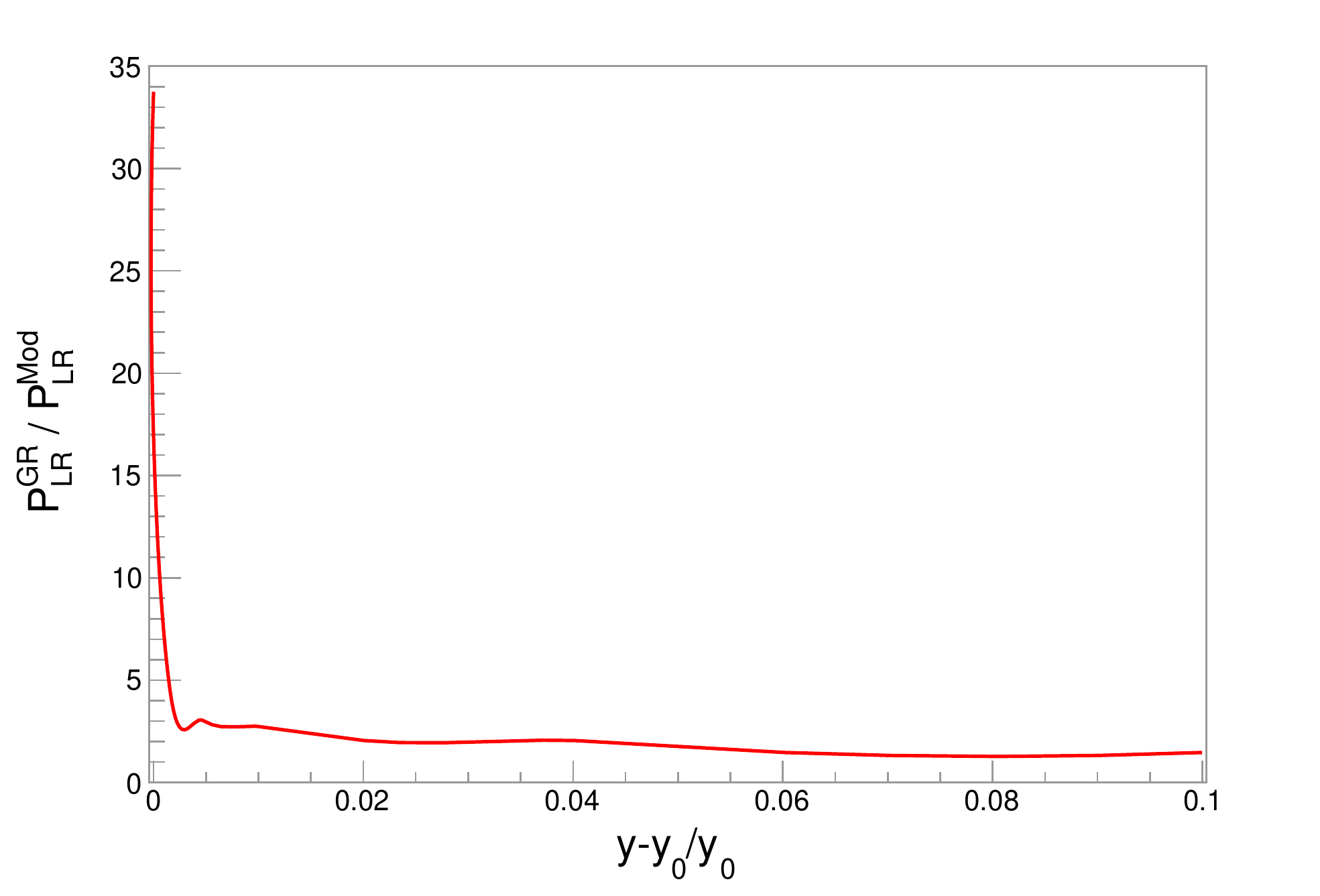}
    \caption{Ratio between the GR probability of spin flip and that of the conformal theory with $A=1.1$ plotted respect to the value of the variable $y$. We have used the value of $\gamma=10$.}
    \label{ConfrontoA}
\end{figure}

\begin{figure}
    \centering
    \includegraphics[scale=0.6]{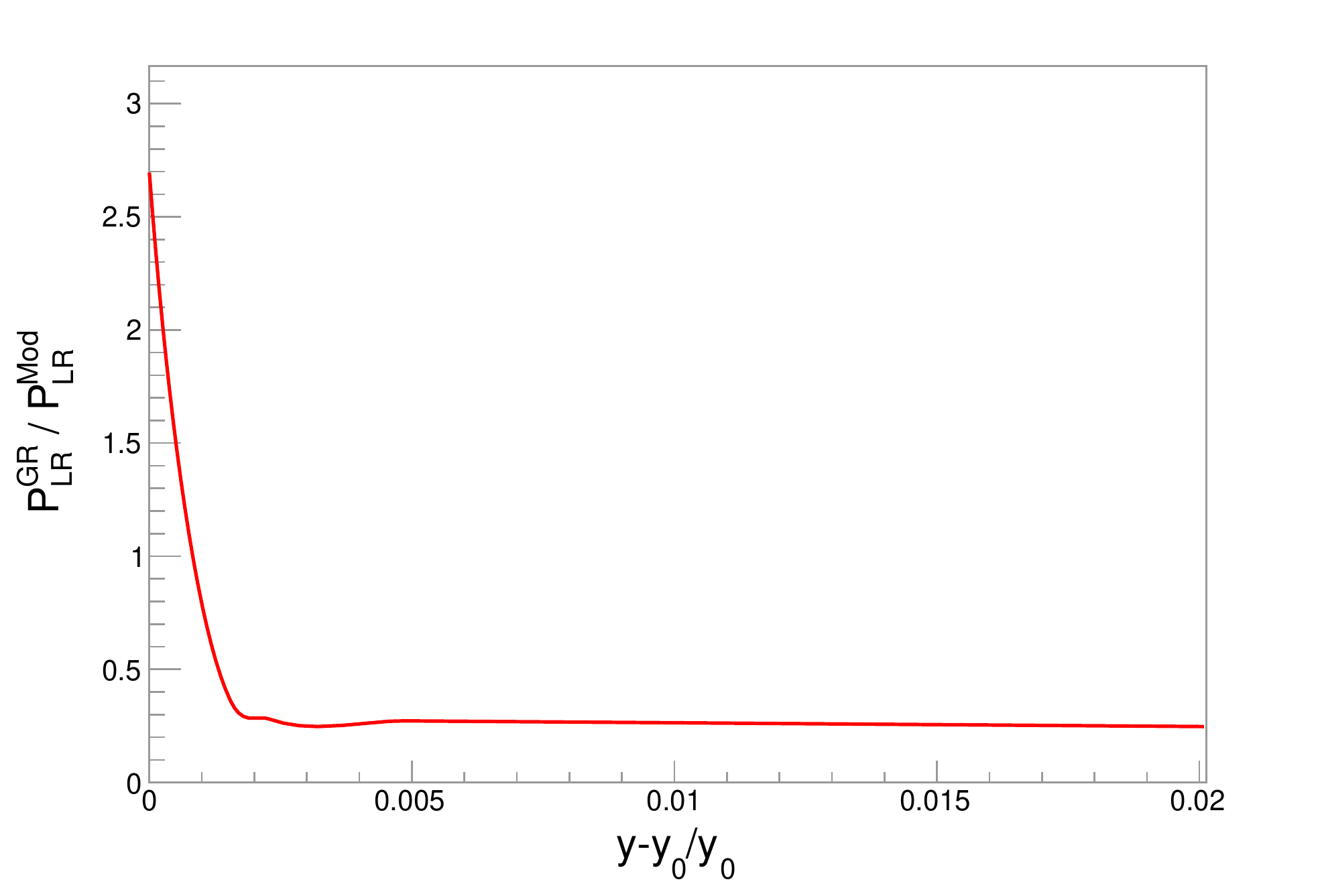}
    \caption{Ratio between the GR probability of spin flip and that of the quintessence theory plotted respect to the value of the variable $y$. We have used the value of $\gamma=10$, $c=0.4$ and $\omega_q=-0.4$.}
    \label{ConfrontoB}
\end{figure}

This analysis turns out relevant in the optic of the recent observations of the event horizon silhouette of a supermassive BH \cite{Aki19}. Indeed, the accretion disk surrounding a BH is a source both of photons (which form, as well known, its visible image) and neutrinos \cite{KawMin07}. The latter suffer gravitational lensing as well as a spin precession in strong external fields near the supermassive BH (it is worth mentioning that
similar effects occur also in a supernova explosion in our galaxy \cite{MenMocQui06}). It is then expected that spin oscillations of these neutrinos
modify the neutrino flux observed in a neutrino telescope.
We have compared the transition probability in presence of a scalar field that screens the gravitational field, and the quintessence field. In both cases, we observe a modification of the spin-flip probability with respect to GR, and, as a consequence, the neutrino fluxes accounting for the interaction with an accretion disk get modified. Results are resumed in Fig. \ref{ConfrontoA} and \ref{ConfrontoB}).
Modification of GR through conformally coupling model (hence screening effects) or quintessence can relevantly affect these astrophysical phenomena, allowing the possibility to find,  with future observations, deviations from GR. 
In the case of the quintessence field, we have also shown the influence of such a field on the nucleosynthesis processes. 

Some final comments are in order: 1)  We have only considered the spin-flip transitions induced by gravitational fields. A more complete analysis requires the inclusion of magnetic field, as in \cite{Dvornikov:2020oay}; 2) The spin flip probability (\ref{PLR18}), when applied to solar neutrinos, gives a probability below the upper bound $0.07$, obtained from the Kamiokande-II \cite{Duan:1992av}.
3) We have assumed that the conformal factor $A(\phi)$ is universal so that the equivalence principle holds \cite{Khoury:2003rn}. However, in the more general case, one may have different $A_i(\phi)$, corresponding to different matter fields $\Psi_i$, with interesting consequences on the spin-flip and spin-flavor oscillations, as well as to spin state abundances of relic neutrinos. All these possibilities will be treated elsewhere.


\begin{acknowledgements}
The work of G.L. and L.M. is supported by the Italian Istituto Nazionale di Fisica Nucleare (INFN) through the ``QGSKY'' project and by Ministero dell'Istruzione, Universit\`a e Ricerca (MIUR).
The computational work has been executed on the IT resources of the ReCaS-Bari data center, which have been made available by two projects financed by the MIUR (Italian Ministry for Education, University and Re-search) in the "PON Ricerca e Competitività 2007-2013" Program: ReCaS (Azione I - Interventi di rafforzamento strutturale, PONa3\_00052, Avviso 254/Ric) and PRISMA (Asse II - Sostegno all'innovazione, PON04a2A)
\end{acknowledgements}

\appendix
\section{Chameleon mechanism}
In some model of scalar-tensor dark energy models, the quintessence is interacting with matter through a conformal coupling scalar field. This coupling may give rise to the screening effect as was studied in the chameleon model. The chameleon model is specified by a power-law potential~\cite{Sadjadi:2020ozc}:
\begin{equation}
    V(\phi)=M^{4+n}\phi^{-n} \,\ ,
\end{equation}
with $n$ positive number, $M$ parameter of mass scale. The effective potential depends on the mass density and the equation of motion is given by
\begin{equation}
    \Box\phi=V_{,\phi}-A^3(\phi)A_{,\phi}(\phi)\title{g}^{\mu\nu}\tilde{T}_{\mu\nu} \,\ ,
\end{equation}
where $\tilde{T}_{\mu\nu}$ is the energy momentum tensor. Using the relation $g^{\mu\nu}\tilde{T}_{\mu\nu}=\tilde{T}=\tilde{\rho}=-A^{-3}(\phi)\rho$, one gets
\begin{equation}
    \Box\phi=V_{,\phi}-A_{,\phi}(\phi)\rho\,.
    \label{equation potential}
\end{equation}
The effective mass of the field is might be defined using the potential
\begin{align}
    m^2_{\mathrm{min}}&=\frac{\partial^2V_{\mathrm{eff}}}{\partial\phi\partial \phi}\Bigg|_{\phi=\phi_{\mathrm{min}}}\\
    &=V_{,\phi\phi}(\phi_{\mathrm{min}})+\frac{\beta^2\rho}{M^2_p}e^{\beta\phi_{\mathrm{min}}/M_p} \,.
\end{align}
One can then obtain the solution for the field solving Eq.~(\ref{equation potential}) with the boundary condition
\begin{align}
    \frac{d\phi}{dr}=0\quad &\mathrm{at}\quad r\rightarrow 0 \,, \\
    \phi\rightarrow\phi_0\quad &\mathrm{at}\quad r\rightarrow\infty \,.
\end{align}
The solution can be obtained expanding the field as $\phi(r)=\phi_0+\delta\phi$, where $\phi_0$ is the uniform background and $\partial\phi$ is the perturbation induced by the spherical symmetric body (like BH or NS). It turns to be
\begin{equation}
    \frac{d^2\delta\phi}{dr^2}+\frac{2}{r}\frac{d\delta\phi}{dr}=m^2_{\mathrm{min}}(\phi)\delta\phi+\frac{\beta\phi_0}{M_p}\rho(r) \,\ .
\end{equation}
Finally, for a gravitational source of radius $R_{\odot}$, one infers
\begin{equation}
    \phi(R)=\phi_0+\delta\phi_{\mathrm{in}}\frac{1}{R}e^{-m_{\mathrm{min}}R_{\odot}(R-1)} \,.
\end{equation}
where $\delta\phi_{\mathrm{in}}$ is the value of the field at the surface of the body and $R=r/R_{\odot}$. For each $\beta$ constant, this procedure gives the resulted $\phi_0$ as shown in Fig.10 of Ref~\cite{Sadjadi:2020ozc}.

\bibliographystyle{unsrtnat}      
\bibliography{template-PRD.bib}

\end{document}